\documentclass[smallextended]{svjour3}
\usepackage{latexsym,amsmath,amscd,amssymb,graphicx,amsbsy,color,xcolor,amsfonts,mathptmx}
\begin{document}

\title{Uniform Acceleration in General Relativity}
\author{Yaakov Friedman \and Tzvi Scarr}
\institute{Yaakov Friedman \at Jerusalem College of Technology, Departments of Mathematics and Physics, P.O.B. 16031 Jerusalem 91160, ISRAEL
   \\ Tel.: +972-2-675-1184, Fax: +972-2-675-1045  \\ \email{friedman@jct.ac.il}
     \and Tzvi Scarr \at Jerusalem College of Technology, Department of Mathematics, P.O.B. 16031 Jerusalem 91160, ISRAEL
     \\ Tel.: +972-2-675-1274, Fax: +972-2-675-1285  \\ \email{scarr@g.jct.ac.il}}
\date{Received: date / Accepted: date}

\maketitle

\begin{abstract}
We extend de la Fuente and Romero's \cite{Romero} defining equation for \emph{uniform acceleration in a general curved spacetime} from linear acceleration to the full Lorentz covariant uniform acceleration. In a flat spacetime background, we have explicit solutions. We use generalized Fermi-Walker transport to parallel transport the Frenet basis along the trajectory. In flat spacetime, we obtain velocity and acceleration transformations from a uniformly accelerated system to an inertial system. We obtain the time dilation between accelerated clocks. We apply our acceleration transformations to the motion of a charged particle in a constant electromagnetic field and recover the \emph{Lorentz-Abraham-Dirac} equation.
\vskip0.2cm


\end{abstract}


\section{Introduction}\label{Intro}
$\;\;\;$

In \cite{Romero}, the authors present a differential equation which characterizes \emph{uniform rectilinear acceleration}, or \emph{hyperbolic motion}, in a general curved spacetime. An accelerometer carried by such a rectilinearly uniformly accelerated observer shows a \emph{constant} linear acceleration. Thus, their equation satisfies Einstein's criterion of ``constant acceleration in the instantaneously co-moving inertial frame." In \cite{FS1}, however, we showed that, besides hyperbolic motion, there are three additional classes of uniformly accelerated motion, including rotational acceleration with constant angular velocity. In this paper, we introduce a differential equation which characterizes \emph{all four classes} of uniform acceleration in a general curved spacetime.

We note that hyperbolic motion itself is \emph{not} Lorentz covariant (see \cite{FS1}). To obtain covariance, linear and rotational accelerations must be handled \emph{together}. Thus, ``constant acceleration in the instantaneously co-moving inertial frame" means not only that an accelerometer carried by the observer shows a constant linear acceleration, but also that his gyroscope shows a constant angular velocity. Note that these are locally measurable quantities.

We will show that in a flat background, the equation introduced here corresponds to our previous definition (\cite{FS1} and \cite{FS2}).  In \cite{FS2}, working in a flat background, we also introduced defining equations for a uniformly accelerated \emph{frame} and established that all four classes of uniform acceleration do, in fact, have constant acceleration in the comoving frame.  However, until now, it was still an open question whether there were more classes of uniformly accelerated motion which were not modelled by our equation. In the current paper, we settle this issue by showing that, in a flat background, every uniformly accelerated motion satisfies our equation. Thus, our equation provides a complete and covariant description of uniformly accelerated motion in flat spacetime.  Moreover, we show here that our theory extends naturally to curved spacetimes.

In this paper, we further develop the theory of covariant uniform acceleration in flat spacetime by deriving velocity and acceleration transformations from a uniformly accelerated frame to an inertial frame. We also derive here the time dilation between clocks in a uniformly accelerated frame.

The plan of this paper is as follows. In section \ref{defua}, we construct the Frenet-Serret frame attached to an observer and present the defining equation for uniformly accelerated motion in a general curved spacetime. The definition uses generalized Fermi-Walker transport. In section \ref{fs}, we show that in flat spacetime, our definition reduces to the one in \cite{FS1,FS2}. Here, we prove one of our main results: in flat spacetime, a motion is uniformly accelerated if and only if it satisfies our defining equation. The remainder of the paper develops the theory of uniform acceleration in a flat background. In section \ref{veltransandtd}, we adapt Horwitz and Piron's notion of ``off-shell" \cite{offshell} to the four-velocity. We call the new notion the \emph{4D velocity} and use it to derive velocity transformations from $K'$ to $K$. We show that when $K'$ is inertial, our velocity transformations reduce to the usual Einstein velocity addition. In section \ref{timedilation}, we show that the normalization factor between the four-velocity and the 4D velocity can be interpreted as a \emph{time dilation} factor.  This leads immediately to a general formula for the time dilation between clocks located at different positions in $K'$. We call the time dilation factor $\tilde{\gamma}$ and show that it is a generalization of the usual $\gamma$ Lorentz factor. We also derive here a formula for the \emph{relativistic angular velocity} of a uniformly accelerated body. In section \ref{arbobs}, we show that an arbitrary rest point of a uniformly accelerated frame is also uniformly accelerated. However, the acceleration tensor gets multiplied by $\tilde{\gamma}$, the time dilation factor. We obtain an explicit expression for the four-acceleration in $K$ of a point at rest in $K'$ and explain the physical meaning of all the terms of this expression. In section \ref{acctrans}, an explicit expression for the four-acceleration in $K$ of an \emph{arbitrarily} accelerating particle in $K'$ is derived. This transformation depends on the position, velocity, and acceleration of the particle in $K'$ as well as the acceleration of $K'$ with respect to $K$. Also here, we explain the physical meaning of all the terms. We apply our acceleration transformations to the motion of a charged particle in a constant electromagnetic field and recover the \emph{Lorentz-Abraham-Dirac} equation. All of our formulas have the appropriate classical limit.

\section{The Definition of Uniform Acceleration}\label{defua}

\subsection{Notation}\label{notation}
$\;\;\;$
Consider a time-orientable four-dimensional differential manifold $\cal{M}$ endowed with a metric $g_{\mu\nu}$ of Lorentzian signature $(+,-,-,-)$. A tangent vector $v$ at a given point of $M$ is \emph{timelike} if $g(v,v)>0$, \emph{spacelike} if $g(v,v)<0$, and \emph{null} if $g(v,v)=0$. Let $\gamma:I\to {\cal M}$, $I$ an open interval of ${\mathbb R}$, be a smooth future-pointing timelike curve, parameterized by the arclength $ds=\sqrt{g_{\mu\nu}dx^\mu dx^\nu}$. In a local system of coordinates $x^\mu$, the curve $\gamma(s)$ is described by a set of four functions $x^\mu(s)$. We assume that the metric satisfies the metricity condition $dg_{\mu\nu}/{ds}=0$.

At every point of $\gamma$, the \emph{four-velocity} $u^\mu$ is defined by
\begin{equation}\label{4vel}
u^\mu=\frac{dx^\mu}{ds}
\end{equation}
and has unit length:
\begin{equation}\label{vel-sq}
u^2=g_{\mu\nu}u^\mu u^\nu=1.
\end{equation}

The \emph{covariant derivative} $\frac{D}{ds}$ of a four-vector $w^\mu$ along $\gamma(s)$ is defined \cite{Hehl} by
\begin{equation}\label{absder}
\frac{Dw^\mu}{ds} = \frac{dw^\mu}{ds} + \Gamma^\mu_{\;\;\alpha\nu}w^\alpha u^\nu,
\end{equation}
where the Christoffel symbols $\Gamma^\mu_{\;\;\alpha\nu}$ are computed from the metric $g$. The \emph{four-acceleration} $a^\mu$ is the covariant derivative of the four-velocity:
\begin{equation}\label{acc}
a^\mu=\frac{Du^\mu}{ds} = \frac{du^\mu}{ds} + \Gamma^\mu_{\;\;\alpha\nu}u^\alpha u^\nu.
\end{equation}
Differentiating $u^2=1$, we have
\begin{equation}\label{ort}
\frac{Du}{ds}\cdot u=0,
\end{equation}
meaning that the four-acceleration is orthogonal to the four-velocity. Since the four-velocity is timelike, the four-acceleration is spacelike.

The \emph{plane of simultaneity} or the \emph{restspace} at the point $\gamma(s)$ is the linear subspace of four-vectors $w$ such that $w\cdot u(s)=0$.

\subsection{Frenet-Serret Basis}\label{fb}
$\;\;\;$
We now construct an orthonormal basis $\{\lambda_{(0)},\lambda_{(1)},\lambda_{(2)},\lambda_{(3)}\}$, $\lambda_{(\alpha)}=\lambda_{(\alpha)}(s)$, of the tangent space at the point $\gamma(s)$. The orthonormality condition means that
\begin{equation}\label{on1}
\lambda_{(\alpha)}\cdot\lambda_{(\beta)}=\eta_{\alpha\beta},
\end{equation}
where $\eta_{\alpha\beta}$ is the Minkowski metric $\operatorname{diag}(1,-1,-1,-1)$. For ease of notation, we represent covariant differentiation of $\lambda_{(\alpha)}$ by
$\dot{\lambda}_{(\alpha)}$. Differentiating (\ref{on1}), we obtain
\begin{equation}\label{on2}
\dot{\lambda}_{(\alpha)}\cdot\lambda_{(\beta)}=-\lambda_{(\alpha)}\cdot\dot{\lambda}_{(\beta)}.
\end{equation}
In particular,
\begin{equation}\label{on3}
\lambda_{(\alpha)}\cdot\dot{\lambda}_{(\alpha)}=0.
\end{equation}

First, let
$\lambda_{(0)}(s)=u(s)$. The four-acceleration $a=\dot{u}$ is spacelike and orthogonal to the four-velocity $u$. We assume that $\dot{u}\neq 0$ for all $s$. Set $\kappa=\sqrt{-a^2}$ and define
\begin{equation}\label{deflam1}
\lambda_{(1)}(s)=\frac{a(s)}{\kappa}.
\end{equation}
The unit vector $\lambda_{(1)}(s)$ gives the direction of the four-acceleration. The scalar $\kappa(s)$ is the magnitude of the four-acceleration and is also called the curve's \emph{curvature}.
From (\ref{deflam1}), we trivially get the first Frenet equation
\begin{equation}\label{dotlam0}
\dot{\lambda}_{(0)}=\kappa\lambda_{(1)}.
\end{equation}
Using the Gram-Schmidt procedure, (\ref{on1}) and (\ref{on2}), we construct a vector $v_{(2)}$ which is orthogonal to both $\lambda_{(0)}$ and $\lambda_{(1)}$:
\begin{equation}\label{defv2}
v_{(2)}=\dot{\lambda}_{(1)}-(\dot{\lambda}_{(1)}\cdot\lambda_{(0)})\lambda_{(0)}=\dot{\lambda}_{(1)}-\kappa\lambda_{(0)}.
\end{equation}
Let $\tau_1=\sqrt{-(v_{(2)})^2}>0$, and define
\begin{equation}\label{deflam2}
\lambda_{(2)}=\frac{v_{(2)}}{\tau_1}.
\end{equation}
Then, from (\ref{defv2}), we have
\begin{equation}\label{dotlam1}
\dot{\lambda}_{(1)}=\kappa\lambda_{(0)}+\tau_1\lambda_{(2)}.
\end{equation}
Similarly, we construct a vector $v_{(3)}$ orthogonal to $\lambda_{(0)},\lambda_{(1)}$ and $\lambda_{(2)}$:
\begin{equation}\label{defv3}
v_{(3)}=\dot{\lambda}_{(2)}-(\dot{\lambda}_{(2)}\cdot\lambda_{(0)})\lambda_{(0)}+(\dot{\lambda}_{(2)}\cdot\lambda_{(1)})\lambda_{(1)}.
\end{equation}
Now (\ref{on2}), (\ref{dotlam0}) and (\ref{on1}) imply that $\dot{\lambda}_{(2)}\cdot\lambda_{(0)}=0$, while (\ref{on2}), (\ref{dotlam1}) and (\ref{on1}) imply that
$\dot{\lambda}_{(2)}\cdot\lambda_{(1)}=\tau_1$. Let $\tau_2=\sqrt{-(v_{(3)})^2}>0$, and define
\begin{equation}\label{deflam3}
\lambda_{(3)}=\frac{v_{(3)}}{\tau_2}.
\end{equation}
From (\ref{defv3}) we now obtain
\begin{equation}\label{dotlam2}
\dot{\lambda}_{(2)}=-\tau_1\lambda_{(1)}+\tau_2\lambda_{(3)}.
\end{equation}
Finally, using (\ref{dotlam0}), (\ref{dotlam1}), (\ref{on2}) and (\ref{on3}), we get that $\dot{\lambda}_{(3)}$is parallel to $\lambda_{(2)}$. Then, using (\ref{on2}), (\ref{dotlam2}) and (\ref{on1}), we get $\lambda_{(2)}\cdot\dot{\lambda}_{(3)}=\tau_2$. Hence,
\begin{equation}\label{dotlam3}
\dot{\lambda}_{(3)}=-\tau_2\lambda_{(2)}.
\end{equation}

Let $\Lambda(s)$ be the $4\times 4$ matrix whose $i$th column consists of the components of the vector $\lambda_{(i)}(s)$ in the local basis. Set
\begin{equation}\label{defineA}
A(s)=\left( \begin{array}{cccc}
0 & \kappa(s) & 0 & 0 \\
\kappa(s) & 0 & -\tau_1(s) & 0 \\
0 & \tau_1(s) & 0 & -\tau_2(s) \\
0 & 0 & \tau_2(s) & 0  \end{array} \right).
\end{equation}
 Then equations (\ref{dotlam0}), (\ref{dotlam1}), (\ref{dotlam2}) and (\ref{dotlam3}), taken together, are equivalent to
\begin{equation}\label{serfre}
\frac{D\Lambda(s)}{ds}=\Lambda(s)A(s).
\end{equation}
The physical meaning of $\kappa(s)$ and $\tau_1(s),\tau_2(s)$ is as follows. An observer on $\gamma(s)$ experiences \emph{linear acceleration} of magnitude $\kappa(s)$ in the direction of $\lambda_{(1)}$. The torsion is defined by a 3D vector $\boldsymbol{\omega}=-\tau_2\lambda_{(1)}-\tau_1\lambda_{(3)}$, which is the axis of the observer's rotational acceleration. The magnitude of the rotational acceleration is $\sqrt{\tau_1^2+\tau_2^2}$. Thus, we refer to $A$ as the \emph{acceleration matrix}. Note that $A$ satisfies equation (\ref{serfre}).

\subsection{Defining Uniform Acceleration}\label{defining}
$\;\;\;$
Einstein's intuitive definition of \emph{uniform acceleration} is ``constant acceleration in the instantaneously co-moving inertial frame." Mathematically, this means that both the linear and rotational acceleration maintain the same magnitude and direction. Hence, the  acceleration matrix $A(s)$ of formula (\ref{defineA}) must be constant and nonzero: $A(s)=A$. In addition, the Frenet basis vectors $\lambda_{(\kappa)}$ must be parallel transported in a way consistent with equation (\ref{serfre}). This leads to the following
\begin{definition}\label{defineunifacc}
A timelike curve $\gamma(s)$ represents \emph{uniformly accelerated motion} if the acceleration matrix $A(s)$ corresponding to the Frenet frame $\Lambda(s)$ is constant and nonzero.
\end{definition}

If $\tau_1=0$, then equations (\ref{dotlam0}) and (\ref{dotlam1}) imply that
\begin{equation}\label{rom}
\ddot{\lambda}_{(0)}=\kappa^2\lambda_{(0)},
\end{equation}
which is equivalent to \cite{Romero}, equation (6). This equation describes hyperbolic motion. However, to obtain a fully Lorentz covariant theory, we cannot assume that $\tau_1=0$. This is because hyperbolic motion itself is not covariant, as shown in \cite{FS1}. Note that equation (\ref{rom}) is third order. For a unique solution, one need only specify the initial position, the initial four-velocity, and the initial linear acceleration.

We now show how to parallel transport the Frenet basis vectors $\lambda_{(\kappa)}$ in a way consistent with definition \ref{defineunifacc}. Recall that a four-vector $w^\mu$ is said to be \emph{parallel transported} along $\gamma(s)$ by the Levi-Civita connection if
\begin{equation}\label{pt}
\frac{Dw^\mu}{ds}= 0.
\end{equation}
If the four-acceleration $a(s)=\frac{Du(s)}{ds}$ is nonzero, then the four-velocity $u$ is \emph{not} parallel transported along $\gamma(s)$ by the Levi-Civita connection. This means that the Frenet basis is not parallel transported. The covariant derivative along the curve does not preserve the restspaces of an accelerating particle.

A more appropriate transport is defined using the \emph{generalized Fermi-Walker derivative} ($GFW$) $\frac{\widehat{D} w^\mu}{ds}$ of a four vector $w^\mu$ (see \cite{Hehl}, \cite{Romero}):
\begin{equation}\label{genFWder}
\frac{\widehat{D} w^\mu}{ds}=\frac{D w^\mu}{ds} - \Omega^{\mu}_{\;\;\nu}w^\nu,
\end{equation}
where $\Omega_{\mu\nu}$ is a rank 2 tensor, defined along $\gamma(s)$. Hehl \cite{Hehl} shows that the metric compatibility condition, which follows from the transport of the orthonormal basis, implies that $\Omega_{\mu\nu}$ is antisymmetric.

In order to insure parallel transport of the Frenet basis vectors $\lambda_{(\kappa)}$, we choose
\begin{equation}\label{Atilde}
\Omega(s)=\Lambda(s)A\Lambda^{-1}(s).
\end{equation}
$\Omega(s)$ is antisymmetric with the same Lorentz invariants as $A$. In fact, $\Omega(s)$ is the acceleration matrix $A$ computed in the initial comoving frame. Hence, we refer to $\Omega(s)$ as the \emph{pullback} of the acceleration matrix $A$ to the initial comoving frame. While $A$ is constant along the world line, $\Omega$ varies with $s$.

Multiplying the right side of (\ref{serfre}) by $\Lambda^{-1}(s)\Lambda(s)$, we obtain
\begin{equation}\label{serfre2}
\frac{D\Lambda(s)}{ds}=\Omega(s)\Lambda(s).
\end{equation}
Clearly, the Frenet basis is $GFW$ \emph{parallel transported}. Using (\ref{serfre2}) and (\ref{genFWder}), we have
\begin{equation}\label{FW}
\frac{\widehat{D}\lambda_{(\kappa)}(s)}{ds}=\frac{D \lambda_{(\kappa)}(s)}{ds} - \Omega(s)\lambda_{(\kappa)}(s)=0.
\end{equation}
In equation (\ref{serfre}), the matrix $\Lambda$ is multiplied by $A$ on the \emph{right}. This means that the time evolution of each basis vector depends on \emph{all} of the basis vectors. However, to have parallel transport, each basis vector must be transported without referring to the other basis vectors. This explains why, in equation (\ref{serfre}), $\Lambda$ is muliplied by $\Omega$ on the \emph{left}.

The rate of a uniformly accelerated clock is \emph{constant} in time, since it is exposed to static forces. Therefore, uniformly accelerated observers can synchronize their clocks, and they will remain synchronized for all times. In section \ref{arbobs}, we will show that every rest point of a uniformly accelerated system is also uniformly accelerated. Therefore, any clock at rest in a uniformly accelerated system can be synchronized to any reference clock. In fact, we hypothesize that clock synchronization can be achieved only between uniformly accelerated systems.

\section{Uniform Acceleration in Flat Background}\label{fs}
$\;\;\;$
In this section, we apply the results of the previous section and obtain the equation for uniform acceleration in flat spacetime. To distinguish from the general case, we use the proper time $\tau$ as the parameter along $\gamma$. We also review a few results from \cite{FS2} on the spacetime transformations from a uniformly accelerated frame to an inertial frame and the metric in a uniformly accelerated frame. For details, see \cite{FS1,FS2}.

\subsection{The Equation of Uniform Acceleration}\label{eqnuafs}

In \cite{FS2}, working in flat spacetime, we defined a \emph{frame} to be uniformly accelerated if there are a constant antisymmetric tensor $A_{\mu\nu}$ and a one-parameter family $\{K_\tau\}$ of instantaneously comoving inertial frames, with bases $\{\lambda_{(\kappa)}(\tau):\kappa=0,1,2,3\}$, such that
\begin{equation}\label{uamlam2}
\frac{d\Lambda(\tau)}{d\tau}=A\Lambda(\tau),
\end{equation}
where $\Lambda(\tau)$ is the $4\times 4$ matrix whose $\kappa$th column is the vector $\lambda_{(\kappa)}(\tau)$. This means that $\Lambda(\tau)$ is parallel transported by the generalized Fermi-Walker derivative
\begin{equation}\label{genFWder2}
\frac{\widehat{D} w^\mu}{ds}=\frac{D w^\mu}{ds} - A^{\mu}_{\;\;\nu}w^\nu.
\end{equation}
We then showed that the solutions to (\ref{uamlam2}) have constant acceleration in the comoving frame. We can now prove the converse: in flat spacetime, for every uniformly accelerated motion, there are $A_{\mu\nu}$ and $\Lambda(\tau)$ satisfying (\ref{uamlam2}).

To see this, first note that covariant derivatives becomes normal derivatives, since we are working in flat spacetime. Now, let $u(\tau)$ be the four-velocity of a uniformly accelerated observer. As in section \ref{fb}, construct $\Lambda(\tau)$ and (constant) $A$ such that
\begin{equation}\label{serfreflat}
\frac{d\Lambda(\tau)}{d\tau}=\Lambda(\tau)A.
\end{equation}
By an appropriate change of basis, we may, without loss of generality, assume that $\Lambda(0)=I$. Then, the unique solution to (\ref{serfreflat}) is
\begin{equation}\label{flatsoln}
\Lambda(\tau)=\exp(A\tau)=\left( \sum_{n=0}^{\infty}\frac{A^n}{n!}\tau^n\right).
\end{equation}
Hence, $\Lambda(\tau)A=A\Lambda(\tau)$, and from (\ref{serfreflat}), we obtain (\ref{uamlam2}). Moreover, since $\Lambda$ and $A$ commute, it follows from the definition (\ref{Atilde}) of $\Omega$ that $\Omega(\tau)=A$, and, therefore, the acceleration matrix is constant even in the initial comoving frame.
\vskip1cm
\emph{Therefore, in flat spacetime, a motion is uniformly accelerated if and only if the associated Frenet basis $\Lambda$ and Frenet matrix $A$ satisfy equation (\ref{uamlam2}).}
\vskip1cm
If we choose $\Lambda(0)=I$, equation (\ref{uamlam2}) is equivalent to Mashhoon's \cite{Mash3} frame-defining equation
\begin{equation}\label{Mashhoondeflam2}
 c\frac{d\lambda_{(\kappa)}^{\mu}(\tau)}{d\tau}=A^{(\nu)}_{\;\;(\kappa)}\lambda_{(\nu)}^{\mu}(\tau).
\end{equation}
The unique solution to (\ref{uamlam2}) with arbitrary initial condition $\Lambda(0)$ is
\begin{equation}\label{exponent solution}
\Lambda(\tau)=\exp(A\tau)\Lambda(0).
\end{equation}
The unique solution for $u(\tau)$ can be easily obtained from (\ref{exponent solution}) by noting that $u(\tau)=\lambda_{(0)}(\tau)$. By integrating $u(\tau)$, one obtains the worldline of a uniformly accelerated observer. Similar constructions appear \cite{Mash3} and \cite{mtw}.

In the $1+3$ decomposition of Minkowski space, the acceleration tensor $A$ of equation (\ref{uamlam2}) has the form
\begin{equation}\label{aab}
A_{\mu\nu}(\mathbf{g},\boldsymbol{\omega})=\left(\begin{array}{cc}0 & \mathbf{g}^T\\ &
\\-\mathbf{g}& \Omega \end{array}\right),
\end{equation}
where $\mathbf{g}$ is a 3D vector with physical dimension of acceleration, $\boldsymbol{\omega}$ is a 3D vector with physical dimension of acceleration, the superscript $T$ denotes matrix transposition, and
\[ \Omega = \varepsilon_{ijk}\omega^k, \]
where $\varepsilon_{ijk}$ is the Levi-Civita tensor. The 3D vectors $\mathbf{g}$ and $\boldsymbol{\omega}$ are related to the translational acceleration and the angular velocity, respectively, of a uniformly accelerated motion.

We raise and lower indices using the Minkowski metric $\eta_{\mu\nu}=\operatorname{diag}(1,-1,-1,-1)$. Thus, $A_{\mu\nu}=\eta_{\mu\alpha}A^{\alpha}_{\;\;\nu}$, so
\begin{equation}\label{aab2}
A^{\mu}_{\;\;\nu}(\mathbf{g},\boldsymbol{\omega})=\left(\begin{array}{cc}0 & \mathbf{g}^T\\ &
\\\mathbf{g}&-\Omega \end{array}\right).
\end{equation}
For any four-vector $x=(x^0,\mathbf{x})$, the evaluation of $A$ on $x$ is
\begin{equation}\label{AonX}
   Ax=(\mathbf{g}\cdot\mathbf{x},x^0\mathbf{g}+\boldsymbol{\omega}\times\mathbf{x})\,.
\end{equation}

The equivalence of our approach and Mashhoon's provides us with two ways to differentiate the $\lambda_{(\kappa)}$'s. Let $x^{(\kappa)}$ be a four-vector. Then, using (\ref{Mashhoondeflam2}), we have
\begin{equation}\label{lambdatrick}
x^{(\kappa)}\frac{d\lambda_{(\kappa)}}{d\tau}=x^{(\kappa)}A^{(\nu)}_{\;\;(\kappa)}\lambda_{(\nu)}=A^{(\nu)}_{\;\;(\kappa)}x^{(\kappa)}\lambda_{(\nu)}
=(Ax)^{(\nu)}\lambda_{(\nu)}.
\end{equation}
On the other hand, using (\ref{uamlam2}), we have
\begin{equation}\label{lambdatrick2}
x^{(\kappa)}\frac{d\lambda_{(\kappa)}}{d\tau}=x^{(\kappa)}A\lambda_{(\kappa)}=A\left(x^{(\kappa)}\lambda_{(\kappa)}\right)=Ax\,.
\end{equation}
We will use these methods interchangeably in what follows.

\subsection{Spacetime Transformations and the Metric}\label{stmetfs}
$\;\;\;$

Let $K'$ be a uniformly accelerated frame attached to an observer with worldline $\widehat{x}(\tau)$. Let $\{K_\tau,\Lambda(\tau)\}$ be the corresponding family of instantaneously comoving inertial frames and their bases, with $\Lambda(0)=I$. Then the
\textit{the spacetime transformations} from $K'$, with coordinates $(y^{(0)},y^{(1)},y^{(2)},y^{(3)})=(y^{(0)},\mathbf{y})$, to $K$, with coordinates $x=(x^0,x^1,x^2,x^3)$, are
\begin{equation}\label{xy}
x=\widehat{x}(\tau)+y^{(i)}\lambda_{(i)}(\tau), \;\;\mbox{ with}\;\; \tau=y^{(0)} \;\;\hbox{ and }\;\; i=1,2,3.
\end{equation}

The differential of the transformation (\ref{xy}) at the point $\mathbf{y}$ of $K'$ is
\begin{equation}\label{differential gen2}
dx=\lambda_{(0)}(\tau)
dy^{(0)}+\lambda_{(i)}(\tau)dy^{(i)}+(A\bar{y})^{(\nu)}\lambda_{(\nu)}(\tau)dy^{(0)},
\end{equation}
where $\bar{y}=(0,\mathbf{y})$. In the $1+3$ decomposition (\ref{aab2}), the differential at the point $\mathbf{y}$ of $K'$ is
\begin{equation}\label{differential gen1}
dx=\left((1+\mathbf{g}\cdot\mathbf{y})\lambda_{(0)}+(\boldsymbol{\omega}\times\mathbf{y})^{(i)}
\lambda_{(i)}\right)dy^{(0)}+\lambda_{(j)}dy^{(j)}.
\end{equation}
The metric at the point $\mathbf{y}$ is

\[ s^2=dx^2=\left((1+\mathbf{g}\cdot\mathbf{y})^2-(\boldsymbol{\omega}\times\mathbf{y})^2\right)(dy^{(0)})^2 \]
\begin{equation}\label{metricaty}
-2(\boldsymbol{\omega}\times\mathbf{y})_{(i)}dy^{(0)}dy^{(i)}-\delta_{jk}dy^{(j)}dy^{(k)}.
\end{equation}
This formula was also obtained by Mashhoon \cite{Mash3}. Concerning the metric (\ref{metricaty}), we note that
\begin{itemize}
\item[$\bullet$] it is \emph{static}: it depends only on the \emph{position} in the accelerated frame and not on \emph{time};
\item[$\bullet$] it approaches the Minkowski metric in the classical limit;
\item[$\bullet$] the physical meaning of the terms $\mathbf{g}\cdot\mathbf{y}$ and $\boldsymbol{\omega}\times\mathbf{y}$ will be explained at the end of the next section.
\end{itemize}

\section{Velocity Transformations}\label{veltransandtd}
$\;\;\;$
In this section, we obtain the transformation of a particle's ``off-shell", or \emph{4D velocity}, in a uniformly accelerated frame $K'$ to its four-velocity in the initial comoving inertial frame $K=K_0$. In other words, without loss of generality, we assume that $\Lambda(0)=I$. In the following section, we will compute the time dilation between the clock of a uniformly accelerated observer located at the origin of $K'$ and the clocks at other positions in $K'$.

A particle's four-velocity in $K$ is, by definition, $\frac{dx^\mu}{d\tau_p}$, where $x(\tau_p)$ is the particle's worldline, and $\tau_p$ is the particle's proper time. However, from Special Relativity, it is known that the proper time of a particle depends on its velocity. In addition, it is known that the rate of a clock in an accelerated system also depends on its position, as occurs, for example, for linearly accelerated systems, as a result of gravitational time dilation. Thus, the quantity $d\tau_p$ depends on both the position and the velocity of the particle, that is, on the state of the particle.

Since we do not yet know the particle's proper time, it is not clear how to calculate the particle's four-velocity in $K$ directly from its velocity in $K'$. To get around this problem, we will differentiate the particle's worldline by a parameter $\tilde{\tau}$ instead of $\tau_p$. For convenience, we will choose
$\tilde{\tau}$ to be a constant multiple of the time. For example, we often choose $\tilde{\tau}=\tau$. The same technique was used by Horwitz and Piron \cite{offshell}, using the four-momentum instead of the four-velocity, thereby introducing the area known as``off-shell" electrodynamics.

We now introduce the following definition.
\begin{definition}\label{defn4Dvelocity}
Let $x^\mu(\tilde{\tau})$ be the worldline of a particle with respect to a frame $K$, inertial or non-inertial, parameterized by the parameter $\tilde{\tau}$. The particle's \emph{4D velocity at a point $P$ on the worldline corresponding to the value $\tilde{\tau}_0$ of the parameter} $\tilde{\tau}$ is denoted by $\tilde{u}$ and defined by
\begin{equation}\label{gen4dvelocity}
\tilde{u}^\mu(P)=\frac{dx^\mu}{d\tilde{\tau}}(\tilde{\tau}_0).
\end{equation}
\end{definition}
Note that both the four-velocity and the 3D velocity can be recovered from the 4D velocity. In fact, since the 4D velocity is a tangent vector to the worldline, it is a scalar multiple of the four-velocity. Causality implies that this scalar is positive. Hence, the particle's four-velocity, which is a normalized tangent vector, is
\begin{equation}\label{4veland4dvel}
u=\frac{\tilde{u}}{|\tilde{u}|},
\end{equation}
the normalization of $\tilde{u}$. In the particular case $\tilde{\tau}=t$, where $t$ is the time in an \emph{inertial} frame, then
\begin{equation}\label{inergamma}
|\tilde{u}|=\left|
\frac{dx^\mu}{dt}\right|=\left|(1,\mathbf{v})\right|=\sqrt{1-\mathbf{v}^2}
=\gamma^{-1}.
\end{equation}
This equality can be used to define the $\gamma$ factor. The 3D velocity $\mathbf{v}=\frac{d\mathbf{x}}{d\tilde{\tau}}$ can be recovered from the 4D velocity
$\tilde{u}=(\tilde{u}^0,\tilde{\mathbf{u}})$ as
\begin{equation}\label{3dfrom4d}
\mathbf{v}=\frac{\tilde{\mathbf{u}}}{\tilde{u}^0}.
\end{equation}

We compute now a particle's four-velocity in $K$, given its 4D velocity in $K'$. Let $y^{(\nu)}(\tau)$ be the worldline in the uniformly accelerated frame $K'$ of a moving particle. Let $\tilde{w}^{(\nu)}=\frac{dy^{(\nu)}}{dy^{(0)}}$ denote the particle's 4D velocity in $K'$ with respect to $y^{(0)}=\tau$.
We will calculate the particle's 4D velocity in $K$ at the point $\mathbf{y}$ of $K'$, with respect to $\tau=\gamma^{-1}t$, where $\gamma$
corresponds to the observer's velocity in $K$. Using the differential (\ref{differential gen2}), the particle's 4D velocity in $K$ is
\begin{equation}\label{4velgen1}
 \tilde{u}=\frac{dx}{d\tau}=\frac{dx}{dy^{(0)}} =\frac{dy^{(\nu)}}{dy^{(0)}}\lambda_{(\nu)}(\tau)+(A\bar{y})^{(\nu)}\lambda_{(\nu)}(\tau).
\end{equation}
Thus,
\begin{equation}\label{4velgen12}
\tilde{u}= (\tilde{w}^{(\nu)}+(A\bar{y})^{(\nu)})\lambda_{(\nu)}(\tau).
\end{equation}
The previous formula transforms the 4D velocity in $K'$ to the 4D velocity in $K$. Using (\ref{3dfrom4d}), we can also obtain the transformation of the 3D velocity in $K'$ to the 3D velocity in $K$.

From (\ref{4veland4dvel}), the four-velocity of the particle in $K$ is
\begin{equation}\label{time dilation gen}
u^{(\nu)}\lambda_{(\nu)}(\tau)=\frac{\tilde{u}}{|\tilde{u}|}=\frac{(\tilde{w}^{(\nu)}+
(A\bar{y})^{(\nu)})\lambda_{(\nu)}(\tau)}{|\tilde{w}+A\bar{y}|}.
\end{equation}
Writing $\tilde{w}^{(\nu)}=(1,\mathbf{w})$, the components of the 4D velocity, as defined in (\ref{4velgen12}), with respect to the basis $\lambda(\tau)$ are, in the $1+3$ decomposition,
\begin{equation}\label{4Dvel1plus3}
\tilde{u}^{(\nu)} = \left(1+\mathbf{g}\cdot\mathbf{y},
(\mathbf{w}+\boldsymbol{\omega}\times\mathbf{y})\right).
\end{equation}
The components of the four-velocity are
\begin{equation}\label{4vel1plus3}
u^{(\nu)} = \frac{\left(1+\mathbf{g}\cdot\mathbf{y},
(\mathbf{w}+\boldsymbol{\omega}\times\mathbf{y})\right)}{\sqrt{\left(1+\mathbf{g}\cdot\mathbf{y}\right)^2-\left(\mathbf{w}+
\boldsymbol{\omega}\times\mathbf{y}\right)^2}}.
\end{equation}
In \emph{inertial} frames, all rest points have a common velocity. In an \emph{accelerated} system, the four-velocity of a rest point depends on the point's spatial coordinates and the acceleration of the system. Note that the expression underneath the square root in (\ref{4vel1plus3}) must be nonnegative. This limits the admissible values for $\mathbf{y}$ and is a manifestation of the locality of the spacetime transformations of \cite{FS2}. The same limitation was obtained by Mashhoon \cite{Mash3}. Moreover, by causality, the expression $1+\mathbf{g}\cdot\mathbf{y}$ must be nonnegative.

To understand the meaning of this formula, let
\begin{equation}\label{3DvelinK}
\mathbf{v}_p=\frac{\mathbf{w}+\boldsymbol{\omega}\times\mathbf{y}}
{1+\mathbf{g}\cdot\mathbf{y}}
\end{equation}
denote the 3D velocity of the moving particle. This 3D velocity is composed of the addition of the particle's velocity $\mathbf{w}$ in $K_\tau$ and the rotational velocity at the position of the particle $\mathbf{y}$. These two velocities are added \emph{classically} because they are both in the frame $K_\tau$. As we will show in section \ref{timedilation}, the factor $1+\mathbf{g}\cdot\mathbf{y}$ adjusts for the time dilation between the clock at the origin of $K_\tau$ and the clock at the point $\mathbf{y}$. The limitation $|\mathbf{v}_p|<c$ imposes a limit of the admissible values for $\mathbf{y}$ and is a manifestation of the locality of the spacetime transformations of \cite{FS2}.

In this notation, the components (\ref{4vel1plus3}) of the four-velocity with respect to the basis $\Lambda(\tau)$ are
\begin{equation}\label{4vel1plus3factored}
u^{(\nu)}=\frac{(1,\mathbf{v}_p)}
{\sqrt{1-v_p^2}}=\gamma (v_p)(1,\mathbf{v}_p),
\end{equation}
which is the usual formula for the four-velocity of an object with 3D velocity $\mathbf{v}_p$.  The \emph{relativistic energy} $E$ of the particle can be obtained by multiplying the zero component of (\ref{4vel1plus3factored}) by the rest-mass $m_0$ of the particle. Thus,
\begin{equation}\label{4momentum}
E=\frac{m_0}
{\sqrt{1-v_p^2}},
\end{equation}
with the 3D velocity $\mathbf{v}_p$ defined by (\ref{3DvelinK}).

Suppose $A=0$ or $\bar{y}=0$. Then the basis vectors $\lambda_{(\nu)}$ of $K_\tau$ are the columns of the matrix of the Lorentz transformation $L:K_\tau \rightarrow K$.
From (\ref{time dilation gen}), the four-velocity in $K$ is
\begin{equation}\label{Ais0oryis0}
u^{(\nu)}\lambda_{(\nu)}(\tau)=\frac{\tilde{w}^{(\nu)}}{|\tilde{w}|}\lambda_{(\nu)}(\tau)=L\left(\frac{\tilde{w}}{|\tilde{w}|}\right).
\end{equation}
This implies that, in this case, the four-velocity transformation from $K'$ to $K$ is the usual Einstein velocity addition between inertial frames.
To show this explicitly, suppose that $K_\tau$ is moving with uniform 3D velocity $\mathbf{v}=(v,0,0)$ with respect to the inertial frame $K$. Suppose a particle has 3D velocity $\mathbf{w}$ in $K_\tau$. We wish to compute $\mathbf{v}\oplus_E\mathbf{w}$, defined to be the particle's 3D velocity in $K$. The particle's 4D velocity in $K_\tau$ with respect to the time $t$ in $K_\tau$, is $\tilde{w}^{(\nu)}=(1,\mathbf{w})$. From (\ref{Ais0oryis0}), we get
\[ u=\frac{1}{|\tilde{w}|}\left( \begin{array}{cccc} \gamma & \gamma v & 0 & 0 \\
\gamma v & \gamma &0 & 0 \\ 0 & 0 & 1 & 0 \\ 0 & 0 & 0 & 1\end{array} \right)\left( \begin{array}{c} 1 \\ w^1 \\ w^2 \\ w^3  \end{array} \right) \]
\begin{equation}\label{utildeeva}
=\frac{1}{|\tilde{w}|}\left(\gamma(1+w^1v),\gamma(v+w^1),w^2,w^3  \right).
\end{equation}
Using (\ref{3dfrom4d}), we obtain the 3D velocity of the particle in $K$ as
\begin{equation}\label{eva}
\mathbf{v}\oplus_E\mathbf{w}=\frac{\left(v+w^1,\gamma^{-1}w^2,\gamma^{-1}w^3  \right)}{1+w^1v}=\frac{\mathbf{v}+P_{\mathbf{v}}\mathbf{w}+\gamma^{-1}(I-P_{\mathbf{v}})
\mathbf{w}}{1+\mathbf{w}\cdot\mathbf{v}},
\end{equation}
where $P_{\mathbf{v}}\mathbf{w}$ denotes the projection of $\mathbf{w}$ onto $\mathbf{v}$. This is the well-known Einstein velocity addition formula (see \cite{Rindler}, formula (3.7)).

If $A\neq 0$ and $\bar{y}\neq 0$, then the velocity addition requires the additional term
\begin{equation}\label{4dveld2accel}
\tilde{u}_a=(A\bar{y})^{(\nu)}\lambda_{(\nu)}(\tau),
\end{equation}
whose components in the basis $\lambda(\tau)$ are, in the $1+3$ decomposition,
\begin{equation}\label{ua1plus3}
\tilde{u}_a^{(\nu)} = \left(\mathbf{g}\cdot\mathbf{y} ,
\boldsymbol{\omega}\times\mathbf{y}\right).
\end{equation}
The zero component, when multiplied by $m_0$, yields the particle's \emph{potential energy}
\begin{equation}\label{potenergy}
V(\mathbf{y})=m_0\mathbf{g}\cdot\mathbf{y}
\end{equation}
due to its position at $\mathbf{y}$.
This additional energy causes the gravitational time dilation and is due to the linear acceleration of $K'$. The quantity $\boldsymbol{\omega}\times\mathbf{y}$ is the rotational velocity of the point $\mathbf{y}$ in $K'$.

\section{Time Dilation}\label{timedilation}
$\;$
We turn now to time dilation. Consider a moving particle in the uniformly accelerated frame $K'$, with worldline $y^{(\nu)}(\tau)$ in $K'$. Let $\tilde{w}^{(\nu)}=\frac{dy^{(\nu)}}{dy^{(0)}}$ denote the particle's 4D velocity in $K'$ with respect to $y^{(0)}=\tau$. From (\ref{4velgen12}), the particle's 4D velocity in $K$ is
\begin{equation}\label{4velgen13}
 \tilde{u}= (\tilde{w}^{(\nu)}+(A\bar{y})^{(\nu)})\lambda_{(\nu)}(\tau),
\end{equation}
and its four-velocity in $K$ is
\begin{equation}\label{fourvelinK}
u=\frac{\tilde{u}}{|\tilde{u}|}.
\end{equation}
On the other hand, denoting the proper time of the particle by $\tau_p$ and using the chain rule, we have
\begin{equation}\label{dtauyfromlenu}
u=\frac{dx}{d\tau_p}=\frac{dx}{d\tau}\frac{d\tau}{d\tau_p}=\tilde{u}\frac{d\tau}{d\tau_p}.
\end{equation}
Comparing (\ref{fourvelinK}) and (\ref{dtauyfromlenu}), we arrive at
\begin{equation}\label{dtauyequdtau}
\frac{d\tau}{d\tau_p}=\frac{1}{|\tilde{u}|}.
\end{equation}
Define
\begin{equation}\label{defgamma}
\tilde{\gamma}=\frac{1}{|\tilde{u}|}.
\end{equation}
Thus, the time dilation is
\begin{equation}\label{dtauyequdtau2}
\frac{d\tau}{d\tau_p}=\tilde{\gamma},
\end{equation}
and, from (\ref{4velgen12}), it follows that the time dilation between two clocks in a uniformly accelerated system depends on their relative position and velocity, as well as on the acceleration of the system. By composing two time dilations, it is straightforward to compute the time dilation between \emph{any two uniformly accelerated clocks}.

The definition of $\tilde{\gamma}$ for accelerated systems is analogous to the definition of $\gamma$ for inertial systems. In fact, we will see below that if $A=0$, then $\tilde{\gamma}=\gamma$.
The factor $\tilde{\gamma}$ expresses the time dilation between the clock at rest at $\mathbf{y}$ and the observer's clock at the origin of $K'$. To obtain the time dilation of the particle with respect to the inertial frame $K$, one must also multiply by the time dilation of the observer with respect to $K$. This factor is the zero component of the observer's four-velocity (for explicit formulas, see \cite{FS1}).

We now express the time dilation (\ref{dtauyequdtau}) in the $1+3$ decomposition. If a particle has 4D velocity $\tilde{w}^{(\nu)}=\frac{dy^{(\nu)}}{dy^{(0)}}=(1,\mathbf{w})$ in $K'$, then using (\ref{4Dvel1plus3}), the time dilation between the particle and the observer is given by
\[ d\tau_p=\left(1+\mathbf{g}\cdot\mathbf{y}\right)\sqrt{1-v_p^2}
\,\,d\tau ,\,\, \]
implying that
\begin{equation}\label{gtimedil3}
   \tilde{\gamma}=\frac{1}{\left(1+\mathbf{g}\cdot\mathbf{y}\right)\sqrt{1-v_p^2}}= \frac{\gamma (\mathbf{v}_p)}{1+\mathbf{g}\cdot\mathbf{y}},
\end{equation}
where the 3D velocity $\mathbf{v}_p$ of the particle is defined by (\ref{3DvelinK}).
Thus, the time dilation between the particle and the observer in $K'$ is the product of the gravitational time dilation and an additional time dilation due to the velocity of the particle together with the rotational velocity of the system. A similar formula was obtained in \cite{Ni}. Note that the expression underneath the square root must be nonnegative. This limits the admissible values for $\mathbf{y}$ and is a manifestation of the locality of the spacetime transformations of \cite{FS2}. The same limitation was obtained by Mashhoon \cite{Mash3}.

If $A=0$, then
\[  \tilde{\gamma}=\frac{1}{\sqrt{1 -w^2}}=\gamma(\mathbf{w}),\]
expressing the time dilation due to the velocity of the particle in $K'$, which is an inertial system in this case.

For a clock at rest in $K'$, the particular case $ \boldsymbol{\omega}=0$ gives a time dilation of
$1+\mathbf{g}\cdot\mathbf{y}$, which is the known formula for gravitational
time dilation. If $\mathbf{g}=0$, the time dilation is
$\sqrt{1-\left(\boldsymbol{\omega}\times\mathbf{y}\right)^2}$, which is the time
dilation due to the rotational velocity of a rotating system.

The lower order terms of the expansion of the time dilation of (\ref{gtimedil3}) are
\[ \tilde{\gamma}\approx 1+\mathbf{g}\cdot\mathbf{y}-\frac{1}{2}(\mathbf{w}+\boldsymbol{\omega}
\times\mathbf{y})^2\]
\begin{equation}\label{tdfo}
=1+\mathbf{g}\cdot\mathbf{y}-\frac{1}{2}(\boldsymbol{\omega}
\times\mathbf{y})^2-\frac{1}{2}\mathbf{w}^2
-\mathbf{w}\cdot(\boldsymbol{\omega}\times\mathbf{y}).
\end{equation}
The second term represents the gravitational time dilation. The third and fourth terms are the transversal Doppler shifts due to the rotation of the system and the velocity of the particle, respectively. The last term is new in the setting of flat Minkowski space but was also obtained recently by Gr{\o}n and Braeck (\cite{GB}, equation (29)) in Schwarzschild spacetime.

We now obtain the physical meaning of $\boldsymbol{\omega}$ in the acceleration matrix $A$. From (\ref{3DvelinK}), with $\mathbf{w}=0$, the 3D velocity in the comoving frame $K_\tau$ of a rest point $\mathbf{y}$ is, in the $1+3$ decomposition,
\begin{equation}\label{vely3D}
\mathbf{v}_p=\frac{\boldsymbol{\omega}\times\mathbf{y}}
{1+\mathbf{g}\cdot\mathbf{y}}.
\end{equation}
This formula defines the angular velocity of a uniformly accelerated (rotating) body. Note that for rest points on the axis of rotation, we have $\mathbf{v}_p=0$. Also, if $\mathbf{y}$ belongs to the plane perpendicular to $\mathbf{g}$, then $\mathbf{v}_p=\boldsymbol{\omega}\times\mathbf{y}$, the classical angular velocity. In general, one must measure the angular velocity of each point with respect to a common clock, in this case, the clock at the origin. Then, since each point of the rotating object must have the same period, the classical angular velocity must be multiplied by the time dilation between the clock at the origin and the clock at the point in question.

\section{The Acceleration of Rest Points in a Uniformly Accelerated Frame}\label{arbobs}
$\;\;\;$
In this section, we will show that each rest point $\mathbf{y}$ of $K'$ is uniformly accelerated. However, the value of the acceleration differs from point to point.

Since the property of being uniformly accelerated is covariant, it is enough to show that the rest point $\mathbf{y}$ of $K'$ is uniformly accelerated in the initial comoving frame $K_0$.

Using the chain rule and (\ref{dtauyequdtau2}), this point's four-acceleration is
\begin{equation}\label{gammaAu0}
a=\frac{du}{d\tau_p}=\frac{du}{d\tau}\frac{d\tau}{d\tau_p}= \tilde{\gamma}\frac{du}{d\tau}.
\end{equation}
As can be seen from (\ref{time dilation gen}), the components $u^{(\nu)}$ for a \emph{rest point} do not depend on $\tau$. Hence, using (\ref{lambdatrick2}), we conclude that, in the initial comoving frame $K_0$, the four-acceleration $a(\mathbf{y})$ of the rest point $\mathbf{y}$ is
\begin{equation}\label{gammaAu}
 a(\mathbf{y}) =  \tilde{\gamma}(\mathbf{y})  u^{(\nu)} \frac{d\lambda_{(\nu)}(\tau)}{d\tau}=\tilde{\gamma}(\mathbf{y})  u^{(\nu)}A\lambda_{(\nu)}(\tau)=\tilde{\gamma}(\mathbf{y})Au.
\end{equation}

In the particular case $\boldsymbol{\omega}=0$, we have $\tilde{\gamma}=\frac{1}{1+\mathbf{g}\cdot\mathbf{y}}$. Thus, \emph{in the frame} $K_0$, the four-acceleration $a(\mathbf{y})$ of the rest point $\mathbf{y}$ is
\begin{equation}\label{gy}
a(\mathbf{y}) =\frac{\mathbf{g}}{1+\mathbf{g}\cdot\mathbf{y}}.
\end{equation}
The same formula was obtained by Franklin \cite{Franklin}.

The above shows that each rest point of a uniformly accelerated system may itself be considered as the spatial origin of a uniformly accelerated system $K_y$. This amounts to rechoosing the initial comoving frame as the comoving frame to $\mathbf{y}$ at $\tau=0$. The components of the acceleration tensor $A$ \emph{in the frame $K_y$} may be computed as follows. Let $\Lambda$ be the Lorentz transformation from $K_0$ to $K_y$. Then,
\begin{equation}\label{AyDiffK}
A(\mathbf{y})=\Lambda^{-1} A\Lambda.
\end{equation}

Using (\ref{AonX}) and (\ref{4vel1plus3}), we obtain the $1+3$ decomposition of the four-acceleration (\ref{gammaAu}):
\begin{equation}\label{tildea1plus3rest0}
a^{(\nu)}=\tilde{\gamma}^2\left({\mathbf{g}}\cdot (\boldsymbol{\omega}\times\mathbf{y}),\left(1+\mathbf{g}\cdot\mathbf{y}\right)
 {\mathbf{g}}+ \boldsymbol{\omega}\times (\boldsymbol{\omega}\times\mathbf{y})\right).
\end{equation}
The classical limit of this acceleration is the sum of the 3D linear acceleration $\mathbf{g}$ and the centrifugal acceleration $\boldsymbol{\omega}\times (\boldsymbol{\omega}\times\mathbf{y})$ (see \cite{Arnold}). In relativity, we add the factor $\tilde{\gamma}^2$ to take care of the difference between the observer's time and the proper time of the particle. The linear acceleration is modified by the time dilation and gives the term $\left(1+\mathbf{g}\cdot\mathbf{y}\right){\mathbf{g}}$.  The zero component, if multiplied by the $m_0c$, gives the power
\begin{equation}\label{power}
P=m_0\tilde{\gamma}^2\mathbf{g}\cdot (\boldsymbol{\omega}\times\mathbf{y})
\end{equation}
of the force generating the acceleration. In the classical limit, we have
\begin{equation}\label{powerclass}
P=m_0\mathbf{g}\cdot (\boldsymbol{\omega}\times\mathbf{y}).
\end{equation}

\section{Acceleration Transformations in a Uniformly Accelerated
Frame}\label{acctrans}
$\;\;\;$
Our next goal is to obtain a particle's four-acceleration $a$ in $K$, given its position $\mathbf{y}$, 4D velocity $w=\frac{dy}{d\tau}=(1,\mathbf{w})$ and 4D acceleration $b=\frac{dw}{d\tau}$ in $K_\tau$. First, however, we will calculate the 4D acceleration $\tilde{a}=\frac{d\tilde{u}}{d\tau}$ in $K$, where $\tilde{u}$ is the particle's 4D velocity in $K$. Now
\begin{equation}\label{tildea}
\tilde{a}=\frac{d\tilde{u}}{d\tau}=\frac{d\tilde{u}^{(\nu)}}{d\tau}\lambda_{(\nu)}
+\tilde{u}^{(\nu)}\frac{d\lambda_{(\nu)}}{d\tau}.
\end{equation}
From (\ref{4velgen12}), we have
\begin{equation}\label{new4a2}
\frac{d\tilde{u}^{(\nu)}}{d\tau}=\frac{dw^{(\nu)}}{d\tau}+A\frac{d\bar{y}}{d\tau}=b+A\bar{w},
\end{equation}
where $\bar{w}=(0,\mathbf{w})$. The quantity $d:=b+{A}\bar{w}$ is the acceleration of the particle with respect to the comoving frame $K_\tau$, since this is the part of $\tilde{a}$ which treats $\lambda(\tau)$ as constant.
Using (\ref{lambdatrick2}) and (\ref{4velgen12}), we have
\begin{equation}\label{new4a3}
\tilde{u}^{(\nu)}\frac{d\lambda_{(\nu)}}{d\tau} = A\left(\tilde{u}^{(\nu)}\lambda_{(\nu)}\right)=Aw+A^2\bar{y}.
\end{equation}
Hence,
\begin{equation}\label{atildesof}
\tilde{a}=b+A\bar{w}+Aw+A^2\bar{y}.
\end{equation}

Using (\ref{AonX}), we now write equation (\ref{atildesof}) in the $1+3$ decomposition. Write $b^{(\nu)}=(0,\mathbf{a}_p)$, where $\mathbf{a}_p$ is the 3D acceleration of the particle in $K'$. Then
\begin{equation}\label{tildea1plus3}
\tilde{a}^{(\nu)}=\left(2 {\mathbf{g}}\cdot\mathbf{w}+{\mathbf{g}}\cdot ({\boldsymbol{\omega}}\times\mathbf{y}),\mathbf{a}_p+\left(1+\mathbf{g}\cdot\mathbf{y}\right)
 {\mathbf{g}}+2{\boldsymbol{\omega}}\times\mathbf{w}+ {\boldsymbol{\omega}}\times ({\boldsymbol{\omega}}\times \mathbf{y}) \right).
\end{equation}
Note the additional terms which appear here and not in (\ref{tildea1plus3rest0}). We now have an acceleration $\mathbf{a}_p$ in $K'$, which adds classically. We also have two additional terms due to the velocity $\mathbf{w}$. The term $2{\boldsymbol{\omega}}\times\mathbf{w}$ is the \emph{Coriolis acceleration} (see \cite{Arnold}). There is also an addition of $2m_0\mathbf{g}\cdot\mathbf{w}$ to the power. This addition is due to the change in the potential energy $V(\mathbf{y})$ (see formula (\ref{potenergy})) and the change in the basis of the comoving frame. This explains the factor of $2$ in this term.

In order to compute the particle's four-acceleration in $K$, we first compute $\frac{d\tilde{\gamma}}{d\tau}$. Using (\ref{defgamma}) and $\tilde{\gamma}\tilde{u}=u$, we have
\begin{equation}\label{gammatag}
\frac{d\tilde{\gamma}}{d\tau}=\frac{d}{d\tau}\frac{1}{|\tilde{u}|}=-\frac{1}{|\tilde{u}|^3}\tilde{u}\cdot\tilde{a}=-\tilde{\gamma}^2u\cdot\tilde{a}.
\end{equation}
Thus, the particle's four-acceleration in $K$ is
\begin{equation}\label{4accfrom4daccel}
a=\frac{du}{d\tau_p}=\frac{du}{d\tau}\frac{d\tau}{d\tau_p}=\tilde{\gamma}
\frac{d}{d\tau}\left(\tilde{\gamma}\tilde{u}\right)=\tilde{\gamma}^2\tilde{a}-\tilde{\gamma}^2(\tilde{a}\cdot u)u.
\end{equation}
Writing $d=b+{A}\bar{w}$, the formula (\ref{atildesof}) can be written $\tilde{a}=d+A\tilde{u}$. Also notice that
\[  A\tilde{u}\cdot u =  A\tilde{u}\cdot \tilde{\gamma}\tilde{u}=0,\]
since $A$ is antisymmetric. Thus, formula (\ref{4accfrom4daccel}) becomes
\begin{equation}\label{4accfrom4daccel2}
a=\tilde{\gamma}^2  {A}\tilde{u}+\tilde{\gamma}^2 d-\tilde{\gamma}^2 (d\cdot u)u
=\tilde{\gamma} \left( {A}u+\tilde{\gamma}( d- (d\cdot u)u)\right) .
\end{equation}
Let $P_ud$ be the projection of $d$ onto $u$, and let $d_{\perp}=(I-P_u)d$.
Then we can write the four-acceleration as
\begin{equation}\label{4accperp2}
a=\tilde{\gamma}^2 ( {A}\tilde{u}+d_\perp)=\tilde{\gamma} {A}u+
\tilde{\gamma}^2d_\perp.
\end{equation}

The explanation of formula (\ref{4accperp2}) is as follows. The acceleration of the origin of $K'$ in $K$ is $Au$, where $u$ is the four-velocity of the origin of $K'$ in $K$. The factor $\tilde{\gamma}$ is the time dilation factor between the clock at the origin of $K'$ and the clock at the position of the particle and arises here because we differentiated the four-velocity by $\tau$ instead of $\tau_p$. The term $\tilde{\gamma} {A}u$ accounts only for the acceleration of $K'$ with respect to $K$. Thus, we must add the term $\tilde{\gamma}^2d_\perp$ to account for the acceleration of the particle inside $K'$. However, since the four-acceleration is always perpendicular to the four-velocity, and $ {A}u$ is perpendicular to $u$, the four-acceleration can contain only the component of $d$ which is perpendicular to $u$. This completes the explanation of formula (\ref{4accperp2}).

Consider the motion of a charged particle in a constant electromagnetic field $F$. We decompose its motion into motion under a constant Lorentz force and acceleration produced by the self-force due to the radiation.  We consider the particle to be at the origin ($\tilde{\gamma}=1$) of a uniformly accelerated system $K'$, with acceleration tensor $A=\frac{e}{m}F$, ignoring the self-force. The acceleration due to the radiation will be considered as motion with respect to $K'$.  The self-force generates an acceleration, which is known to be $d=-\tau_0 A^2u$, where $\tau_0$ is a universal constant. Thus, in this case, formula (\ref{4accfrom4daccel2}) coincides with the  Lorentz-Abraham-Dirac equation (\cite{Rohrlich2}, equation S-6, page 258)
\begin{equation}\label{rohreq}
\frac{du}{d\tau}=Au - \tau_0\left(A^2u - (A^2 u \cdot u)u\right),
\end{equation}
which Rohrlich calls the \emph{correct} equation of motion of a classical point charge.

\section{Conclusions and Future Directions}\label{conc}
$\;\;\;$
We have introduced here a new equation (Definition \ref{defineunifacc}) which defines \emph{uniform acceleration} in a general curved spacetime. The solutions to this equation have constant acceleration in the instantaneously co-moving inertial frame, as suggested by Einstein. This improves the results of \cite{Romero}, which handle only hyperbolic motion. In section \ref{fs}, we have explicit solutions in a flat space background.

In a flat background, we have obtained velocity and acceleration transformations (sections \ref{veltransandtd} and \ref{acctrans}). To this end, we introduced here the \emph{4D velocity} (Definition \ref{defn4Dvelocity}, section \ref{veltransandtd}) and the \emph{4D acceleration}. This is an adaptation of Horwitz and Piron's \cite{offshell} concept of ``off-shell" to the four-velocity. For inertial systems, our velocity transformations reduce to the usual Einstein velocity addition (\ref{eva}).

Formula (\ref{gtimedil3}) gives the explicit form of the time dilation between clocks located at different positions in a uniformly accelerated frame. The time dilation is a product of the time dilation due to the potential energy at the position of the clock and the time dilation due to the clock's velocity. The power series expansion (\ref{tdfo}) contains all of the usual terms plus a \emph{new} term which has only been obtained in Schwarzschild spacetime \cite{GB}.

In section \ref{arbobs}, we have shown that every rest point in a uniformly accelerated frame is also uniformly accelerated. However, the value of the acceleration differs from point to point and must be multiplied by the time dilation factor (formula (\ref{gammaAu})).

We consider the case of a charged particle in a constant electromagnetic field. The co-moving frame to this particle is determined by the field, ignoring the self-force. The acceleration caused by the radiation is considered as motion with respect to the co-moving frame. We then apply our acceleration transformations and recover the \emph{Lorentz-Abraham-Dirac} equation.

We want to determine whether the spacetime transformations between uniformly accelerated systems form a \emph{group}. If yes, we want to characterize this group, which will be an extension of the Lorentz group.

We have begun to apply the results here to rotating systems. The case of a disk rotating with constant angular velocity is of particular interest.  The correct form of the spacetime transformations from such a disk to an inertial frame has been debated for over 100 years and continues to the present day. The recent book \cite{rizzi} makes it clear that there is still no universally accepted theory. Applying our methodology, one can obtain explicit spacetime transformations using only the basic tenets of Special Relativity, the inherent symmetries of the problem, and the results of \cite{FS1,FS2} and the current paper. We can also show how to avoid the \emph{time gap} and the \emph{horizon problem}.

Our results may also be applied to the theory known as \textit{Extended Relativity (ER)}, developed in \cite{Fiard}. $ER$ extends Special Relativity by purporting the existence of a \textit{universal maximal acceleration} and has been successfully applied to the hydrogen atom and the harmonic oscillator. Thus far, however, only the one-dimensional case has been treated. We are currently trying to apply the results of \cite{FS1}, \cite{FS2} and the current paper in order to extend $ER$ to full covariance. The first author has performed two experiments, one at the Petra III facility at DESY in Hamburg, and one at the ESRF synchrotron in Grenoble, to prove the existence of a universal maximal acceleration and to measure its value. Further experiments are planned.

\begin{acknowledgements}
We would like to thank L. Horwitz, B. Mashhoon, F. Hehl, Y. Itin, S. Lyle, and {\O}. Gr{\o}n for challenging remarks which have helped to clarify some of the ideas presented here. This work was partially supported  by the German-Israeli Foundation for Scientific Research and Development: GIF No. 1078-107.14/2009.
\end{acknowledgements}

\end{document}